\documentstyle[prb,aps,epsf]{revtex}

\begin{document}
\draft
\def\da{\dagger}

\twocolumn[\hsize\textwidth\columnwidth\hsize\csname@twocolumnfalse%
\endcsname

\title{ Bosonization for Wigner-Jordan-like Transformation :\\
Backscattering and Umklapp-processes on Fictitious Lattice.}

\author{D.N. Aristov $^{1,2}$}

\address{
 $^1$ Laboratoire L{\'e}on Brillouin, CE-Saclay,
91191 Gif-sur-Yvette Cedex, France. }
\address{$^2$ Petersburg Nuclear Physics Institute,
Gatchina, St. Petersburg 188350, Russia}
\date{\today}
\maketitle

\begin{abstract}
We analyze the asymptotic behavior of the exponential form in
the fermionic density operators as the function of ruling parameter $Q$.
In the particular case $Q=\pi$ this exponential associates with the
Wigner-Jordan transformation for  $XY$ spin chain model.
We compare the bosonization approach and the evaluation via the
Toeplitz determinant. The use of Szeg\"o-Kac theorem suggests that
at $Q>\pi/3$ the divergent series for intrinsic logarithm provides
a bosonized solution and faster decaying one,
found as the logarithm's value on another sheet of the complex plane.
The second solution is revealed as umklapp-process
on the fictitious lattice while originates from backscattering
terms in bosonized density. Our finding preserves in a wide range of
fermion filling ratios.
\end{abstract}


]

The method of bosonization proved to be an effective tool for
studying the properties of one-dimensional (1D) systems of interacting
fermions. Describing the fermion system through the bosonic collective
modes, \cite{tomonaga} this technique was initially used for
providing the exact solution of Luttinger model. \cite{Ma-Li}
Recently the bosonization attracted a renewed
interest due to investigation of Quantum Hall Effect \cite{WCNC},
two channel Kondo model \cite{Kondo}
 and High-$T_c$ superconductivity. \cite{Luther}
Essential efforts are made to generalize the method for
higher dimensions. \cite{d>1}

In a recent paper \cite{AG}
the bosonization technique was successfully employed
to study the magnetic properties of the high-$T_c$ -related system.
The 1D model with a spin-charge separation was solved and important
physical consequences were obtained.
The bosonization was used to estimate the (time-dependent)
Wigner-Jordan-like exponential in fermionic on-site density
operators $c_{j}^{\da} c_{j}$.  Namely the quantity of the
interest was the average $\left\langle \exp\ i Q
\sum_{j=1}^{l} c_{j}^{\da} c_{j}^{}\right\rangle$  at arbitrary
value of $Q$. This average should be $2\pi-$periodic in $Q$ and
attain the real value at $Q=\pi$ (the case for Wigner-Jordan
transformation).  These properties are lost after the simple use
of the bosonization technique. Although various {\em ad hoc}
tricks were implemented to cure the problem at $Q=\pi$ (see,
e.g.  \cite{VBS} and references therein),  no systematic bosonic
representation existed.

In an attempt to resolve this puzzle we addressed ourself to the
vast available literature on the subject. The results of this
survey could be outlined in the following way.  The
straightforward application of bosonization, though formally
valid, is seemingly incapable to provide a closed form of the
final answer due to the complexity of intermediate
expressions. The key could be found in comparison of two
counterparts for the same problem : the bosonization by Mattis
and Lieb \cite{Ma-Li} and Luttinger's original solution
\cite{Luttinger} via the Toeplitz determinants. The Szeg\"o-Kac
theorem, used in the approach by Luttinger, dates back
\cite{MPW} to Onsager famous solution of 2D Ising model. The
exhaustive literature on the latter subject suggests some useful
hints about the discussed issue.

Basing on the alternative representation of the considered
average, we make a following major statement.  The
above exponential form in the fermionic density operator
reveals a singularity at some finite value of $Q$.
This is verified by an application of Szeg\"o-Kac theorem and
manifests itself as the appearance of umklapp-processes
contribution on the fictitious lattice (see below).
We provide an explicit expression obeying the above
$2\pi$-periodicity and real valued at $Q=\pi$.

The rest of the paper is organized as follows.
We formulate the simplified version of our problem and solve it by
bosonization method in Section I. We rewrite our quantity of interest
as the Toeplitz determinant and apply the Szeg\"o-Kac theorem in Section
II ; the appearing singularity is discussed here. In Section III the
structure of the desired solution is  suggested and justified.
This solution and its physical consequences are discussed
in the Section IV of the paper.

\section{The model and approach by bosonization}

We consider a gas of free spinless fermions on a lattice with
Hamiltonian

       \begin{eqnarray}
       \label{hamilt}
       H_1 &=& \sum_p ( \cos k_F - \cos p )
       c^{\dag}_{p}c^{}_{p}
       \end{eqnarray}
where Fermi wave-vector $k_F=c\pi$.
Our aim is to evaluate a following quantity :

\begin{eqnarray}
\label{spinop1}
{\cal C}(Q,l) =
 \left\langle \exp\ i Q
 \sum_{j=1}^{l} c_{j}^{\da} c_{j}^{} \right\rangle_{H_1} ,
\end{eqnarray}
where  $ c_{j}^{\da} c_{j}^{}$ corresponds to the on-site
density operator.
The expression of the form (\ref{spinop1})
appeared previously in various studies concerning
the physics in one dimension.
 For the particular case of $Q=\pi$
it is found in the Wigner-Jordan transformation for the spin operators
in 1D XY model.
It can be also associated with the incorporation
of interaction effects in the bosonization approach to Luttinger model.
In our problem (see Appendix \ref{app:connec} for details)
the quantity (\ref{spinop1}) connected the observable spin
susceptibility and easier tractable one on the fictitious lattice. The
parameter $Q$ appeared as a wave-vector belonging to the fictitious
lattice and varied over the whole Brillouin zone, $Q\in [-\pi,\pi)$.

The average (\ref{spinop1}) is easily evaluated by introducing the
boson representation for the fermionic density.
 The main steps of this derivation are outlined below.
First we introduce the density operator :

\begin{eqnarray}
\label{dsty}
\rho(q)= N^{-1} \sum_p c^{\da}_{p} c^{}_{p+q} ,
\end{eqnarray}
where $N$ is the total number of lattice sites.

\begin{eqnarray}
\label{fermionic}
{\cal C}(Q,l)= e^{iQcl} \left\langle \exp\left[ i Q \sum_{q}
\phi(q,l) \rho(q)\right]  \right\rangle ,
\end{eqnarray}

\noindent
and we have defined $\phi (l,q)=(1-e^{i q l})/(1-e^{i q})$ and
$\phi (l,0) =l$.
 The non-fluctuating component
of the density $\langle c_j^\da c_j\rangle = c$
is treated separately and
leads to the prefactor $e^{iQcl}$ in Eq.\  \ref{fermionic}.

 Now we may use a well known
procedure\cite{Lu-Pe} in which the fermion density is split into
contributions from right and left moving fermions.  We write

\begin{eqnarray}
\rho(q)&=&\rho_{+}(q)+\rho_{-}(q)  \nonumber \\
&=& N^{-1} \sum_{p>0}
c^{\da}_{p} c^{}_{p+q}+ N^{-1}\sum_{p<0} c^{\da}_{p}
c^{}_{p+q}, \quad q\neq 0 \label{defrho} \\
\rho(q)&=& N^{-1}( N_- + N_+ ),\quad q = 0
\end {eqnarray}
where $N_+$ ($N_-$) are charge operators for right (left)
movers, describing the fluctuations of the Fermi surface
 \cite{Haldane}

The long wave-length density fluctuations of the
$1\!-\!d$ electron gas may be represented by boson operators
defined by the relations  :

\begin{eqnarray}
\label{bosons}
\begin{array}{l}
\rho_{+}(q)=v(q) \left(\theta(q) b_{q}^{\da}+\theta(-q) b_{-q}^{ }\right)
\\
\rho_{-}(q)=v(q) \left(\theta(q) b_{-q}^{ }+\theta(-q) b_{q}^{\da}\right),
\end{array}
\end{eqnarray}

\noindent where $v(q)=\sqrt{|q|/(2N\pi)}$. \cite{Ma-Li}
In terms of the above operators, the Hamiltonian reads  :
       \begin{eqnarray}
       \label{hamilt2}
       H = v_F\sum_q |q|b_q^{\da} b_q +  v_F\frac\pi N (N_+^2 + N_-^2)
       \end{eqnarray}
where Fermi velocity $ v_F = \sin k_F$.

With these definitions
 we may rewrite Eq.\ \ref{fermionic} in the form :

\begin{eqnarray}
{\cal C}(Q,l )&=& e^{iQcl}
\left\langle \exp i\frac{Ql}N (N_+ + N_-)
\right\rangle    \nonumber \\
&& \times
 \left\langle \exp iQ
\sum_{q} v(q)
\phi(q,l) \left[  b_{q}^{\da}+  b_{-q}^{ }\right]
\right\rangle
\label{defb}
\end{eqnarray}
Let us focus on the case of zero temperature; the finite $T$ are
considered in the Appendix \ref{app:temp}.
 The fermionic vacuum corresponds then to the absence of bosons in
the ground state, in addition $N_\pm =0$.
The evaluation of the above expression yields immediately :

\begin{eqnarray}
\label{corrfermions}
{\cal C}(Q,l)=\exp \left[ iQcl -\frac{Q^2}{2 \pi^2} {\cal
F}(l)\right],
\end{eqnarray}

\noindent where

\begin{eqnarray}
\label{decay}
{\cal F}(l )= \pi^2 \sum_{q} v^{2}(q)
\frac{  1-\cos q(l-l')}{1-\cos q}.
\end{eqnarray}

The integration in Eq. \ref{decay} is cutoff at a
wavevector of order $k_F$ since the density fluctuations of
the electron gas are well defined excitations only in the region
$|q|\lesssim k_F$. \cite{cutoff}

\noindent   Evaluating the above
integral in the logarithmic
approximation we find  :

\begin{eqnarray}
\label{defc}
{\cal C}(Q,l ) \propto
{\cal C}_{boson}(Q,l )=  \exp \left[iQcl
- \frac{Q^2}{2\pi^2} \ln l \right] ,
\end{eqnarray}
This expression indicates
the power law decay of the correlations with $l$, which
is well established fact for the 1D systems. \cite{Solyom}

The result (\ref{defc}) seems however unsatisfactory on two
following reasons. First, the initial expression (\ref{spinop1}) was
invariant under the shift $Q\to Q+2\pi$. This property is obviously
absent in
(\ref{defc}). Second, $e^{i\pi c_j^\da c_j} = \pm 1$ for each $j$,
and one should have the real value of  (\ref{spinop1})  at $Q=\pi$,
meanwhile this feature is also not found in (\ref{defc}).

The inadequacy of the expression (\ref{defc}) for the particular case
$Q=\pi$  was noticed by Luther and Peschel \cite{Lu-Pe}.
They argued that the term  $ \propto e^{-iQcl} $ should be added to
 (\ref{defc})
to get the correct result. The necessity of inclusion of such term
was ascribed to the $2k_F$ (``backscattering'')
processes ( we remind that   $2k_F = 2\pi c$
and $Qc - 2k_F = -Qc$).
Really, one might argue that the representation
(\ref{defrho}) of the electronic
density is incomplete and the low-energy fluctuations with $q\simeq 2k_F$
should be explicitly included into the
notation. \cite{Haldane}
These terms are described in the Appendix \ref{app:2kf} where we show
that the ``$2k_F$'' terms produce very complicated and seemingly intractable
expressions.

A general question arises here.  If $2k_F$ processes
 provide the real value of  ${\cal C}(Q,l )$ at $Q=\pi$, then
why these $2k_F$ terms are less important at other values
of {\em independent} parameter $Q$ where
${\cal C}(Q,l)$ is not real ?
Below we argue that at $Q>\pi/3$ the ``marginal'' contribution
of backscattering breaks the analiticity of
${\cal C}(Q,l)$.

In our problem the parameter $Q$ varies
from 0 to $\pi$. Therefore the ignorance of the above discrepancy might
produce non-trivial consequences.
To resolve this question
we formulate below the alternative way for the evaluation of
 ${\cal C}(Q,l)$ (\ref{spinop1})
and provide the answer which explicitly obeys two above conditions,
the $2\pi$-periodicity in $Q$ and real valued quantity
${\cal C}(Q=\pi,l)$.

\section{Szeg\"o-Kac theorem}

We use a modification of an approach originally due to Lieb,
Schultz and Mattis \cite{LSM}.
It is based on a representation
of the  correlation function ${\cal C}(Q,l )$
of Eq.\ (\ref{fermionic}) as
the Toeplitz   determinant

     \begin{eqnarray}
     \label{toeplitz}
     {\cal C}(Q,l)& =&
     det(M^{(l)})                          \\
     M_{km}^{(l)}& = &\delta_{km} - (1-e^{iQ}) g_{km}.
     \nonumber
     \end{eqnarray}

The $l\!\times\!l$ matrix $M^{(l)}$ is constructed out the fermionic
correlation functions
     \begin{eqnarray}
      g_{km} &=& \langle  c_{k}^{\da} c_{m}^{} \rangle
      = \frac1{2\pi} \int_{-\pi}^\pi dq\, n_q e^{iq(k-m)}
      \nonumber \\
      &=& \frac{\sin(\pi c(k-m) )}{\pi(k-m)}
      \end{eqnarray}
 where the Fermi
function $n_q =1 $ at $|q| < k_F$ and $ n_q = 0$ otherwise.

For completeness we sketch below the derivation of  (\ref{toeplitz}).
We observe that

     \begin{eqnarray}
     e^{i Q c_i^\da c_i} =
      (c_i^\da + c_i )(c_i^\da + e^{i Q } c_i )
     \end{eqnarray}
which is readily verified in the representation diagonalizing
$c_i^\da c_i$. Defining

     \begin{eqnarray}
      A_i = c_i^\da + c_i , \quad
      B_i = c_i^\da + e^{i Q } c_i ,
     \end{eqnarray}
we have
     \begin{eqnarray}
     \label{ab}
     {\cal C}(Q,l)& =&
      \langle A_1 B_1 A_2 \ldots B_{l-1} A_l B_l \rangle
     \end{eqnarray}
Zuber and Itzykson  evaluated the square of
the value (\ref{spinop1}) in a similar problem.\cite{Zu-Itz}
They noticed at $Q = \pi$ one had the possibility to identify the
operators $A_i, B_i$  with a new fermionic (Majorana) field.
This is not so for other values of $Q$ since
$A_i B_i + B_i A_i =  1+ e^{iQ} \neq 0 $, therefore the trick used
in \cite{Zu-Itz} with introducing the second Majorana field is
hardly applicable to our case.

However one can
apply the Wick theorem
which permits one to express the vacuum
expectation value (\ref{ab}) in terms of expectation values of products
of just two operators. \cite{AGD}
This procedure is facilitated if one notices that
     \begin{eqnarray}
      \langle A_i A_j   \rangle =  \langle B_i B_j   \rangle =0,
       \quad i\neq j
%
     \end{eqnarray}
and
     \begin{eqnarray}
      \langle A_i B_j \rangle = \delta_{ij} - (1-e^{iQ}) g_{ij}
     \end{eqnarray}

The most straightforward contribution to (\ref{ab}) is
\[
   \langle A_1 B_1\rangle  \langle A_2 B_2\rangle
   \ldots   \langle A_l B_l \rangle
\]
All other pairings are obtained by permuting the $B$'s among themselves
with  $A$'s fixed. \cite{fnote}
The number of crossings of $A$'s by $B$'s is
always even, hence the sign associated with a given permutation is $(-1)^{p}$
where $p$ is the signature of the permutation $P$ of the $B$'s.
As a result

     \begin{eqnarray}
     \label{wick}
     {\cal C}(Q,l)& =&
      \sum_P  (-1)^{p'}
     \langle A_1 B_{P(1)} \rangle
     \langle  A_2  B_{P(2)} \rangle
       \ldots  \langle  A_l B_{P(l)} \rangle
     \end{eqnarray}
and we arrive to the above Eq.\ (\ref{toeplitz}).

Note that the representation
(\ref{toeplitz}) does not appeal to any particular law
of fermionic dispersion, and the notion of
the Fermi wave-vector only appears in $g_{km}$ (at $T=0$).

Now we use the asymptotic behavior of the
above determinant at large $l$ known as Szeg\"o-Kac theorem
\cite{Gr-Sz,MPW},

\begin{equation}
\label{determinant}
     \lim_{l\to\infty} \left[ det(M^{(l)}) D^{-l}
      \right] = \exp \sum_l^\infty K_l K_{-l} l ,
\end{equation}

\noindent
 where we follow the notation by Luttinger \cite{Luttinger}

     \begin{eqnarray}
     D &=& \exp \int_{-\pi}^\pi \frac{dq}{2\pi} \ln f_q,
     \\
     K_l &=&   \int_{-\pi}^\pi \frac{dq}{ 2\pi} e^{-ilq}
      \ln f_q,
     \label{app:kl}
     \end{eqnarray}
Where
     \begin{eqnarray}
     f_q  &=& \sum_{m=-\infty}^{\infty} e^{iq(m-k)} M^{(\infty)}_{km},
     \end{eqnarray}
in our case  $f_q  = 1- (1-e^{iQ}) n_q $.
Therefore at the
first glance one has $\ln f_q = iQ n_q $. Next, $D = iQc$,
$K_l = iQ g_{l0}$   and simple calculation
returns us back to Eq.\   (\ref{defc}).
It was noticed in \cite{Luttinger} however
that the proof of Eq.(\ref{determinant}) required the convergence
of the series for logarithm

     \begin{eqnarray}
     \ln f_q &=&  -\sum _{n=1}^\infty  \frac{(1-f_q)^n}n
     \end{eqnarray}
In his discussion Luttinger explicitly assumed the convergence of the
above series, associating it with the smallness of the interaction between
electrons \cite{Luttinger} (see also \cite{Ma-Li}).
This argument is not applicable to our case since

     \begin{eqnarray}
     \label{logarithm}
     \ln f_q &=& - n_q  \sum _{n=1}^\infty
       \frac{(-2i\sin(Q/2))^n}n e^{iQn/2}
     \end{eqnarray}
Although one might say that this series converges
to the value $iQ n_q$, it is divergent in absolute
values at $|Q| >\pi/3$. Thus formally it can produce an
arbitrary value as its sum upon the reordering of terms.
The consequences of this observation are elucidated below.

Concluding this section we wish to make some remarks concerning
Eq.(\ref{determinant}).

The attentive reader may notice, that the proof of Szeg\"o-Kac
theorem  \cite{Gr-Sz,MCW} usually demands that the function
$f_q$ is continuous. In the absence  of interaction
this is achieved
at finite temperatures, when the Fermi distribution function is
smeared at the scale $q-k_F\sim T/v_F$. In this latter case however
one expects the exponential tail \cite{AG} as a far asymptote of
 ${\cal C}_Q(l)$ and power-law decay as intermediate regime. The
problem is hence somewhat similar to the problem
of correlations in the scaling limit (uniform asymptote)
in the 2D Ising model near $T_c$. \cite{WMTB}
On the other hand, the sufficient conditions for the
validity of Szeg\"o-Kac theorem are apparently not known. \cite{MCW}
The scrupulous investigation of the arising difficulties  would
 bring us far from our initial task. That is why we tried to get
an insight of the situation by the procedure described in the next
section.

\section{the conjectured solution}

\begin{figure}
\centerline{\epsfxsize=8cm \epsfbox{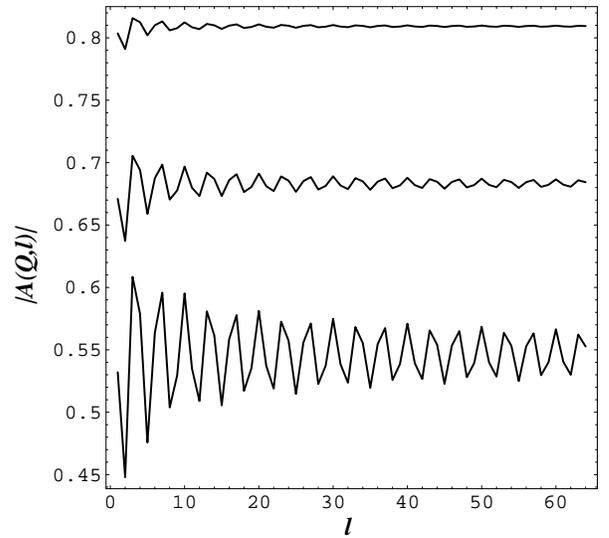}}
\caption{The quantity $|A(Q,l)|$ defined in
Eq.(\ref{def-A}) shown as a function of the chain fragment length $l$.
The plotted values are for $c=0.3$ and for $Q/\pi= 0.45 , 0.6, 0.75 $
from top to bottom, respectively; the lines are connecting the points.
 Note the appearance of faster decaying asymptote at larger $Q$.
\label{fig:ampl}
}
\end{figure}

We numerically  calculated
the determinants  of $l\times l$ matrices of the form
(\ref{toeplitz}) for different values of $c$ and $Q$ and for
$1\leq l\leq L = 64$. We found that the
asymptotic behavior  $det(M^{(l)}) \propto l^{-Q^2/2\pi^2}$ actually
begins with $l$ of order of unity. We illustrate this fact
on the Fig. \ref{fig:ampl} where we plot the auxiliary quantity

     \begin{eqnarray}
     \label{def-A}
      A(Q,l) = det(M^{(l)})\, l\,^{Q^2/2\pi^2}
     \end{eqnarray}
for some particular values of $Q$ and $c$.
If Eq. (\ref{defc}) would hold, $|A(Q,l)|$ would be
constant, which is indeed realized at largest $l$.
An essential additional feature could be also noted on this figure.
Namely the extra oscillating term exists, which decays faster
than the bosonized solution (\ref{defc}) at small $Q$.
However with the increase of $Q$, the decay of this extra term becomes
comparable to one of (\ref{defc}) and both terms are of the same
significance at $Q=\pi$.
This qualitative picture is reproduced at the other values of $c$
as well.

\begin{figure}
\centerline{\epsfxsize=6cm \epsfbox{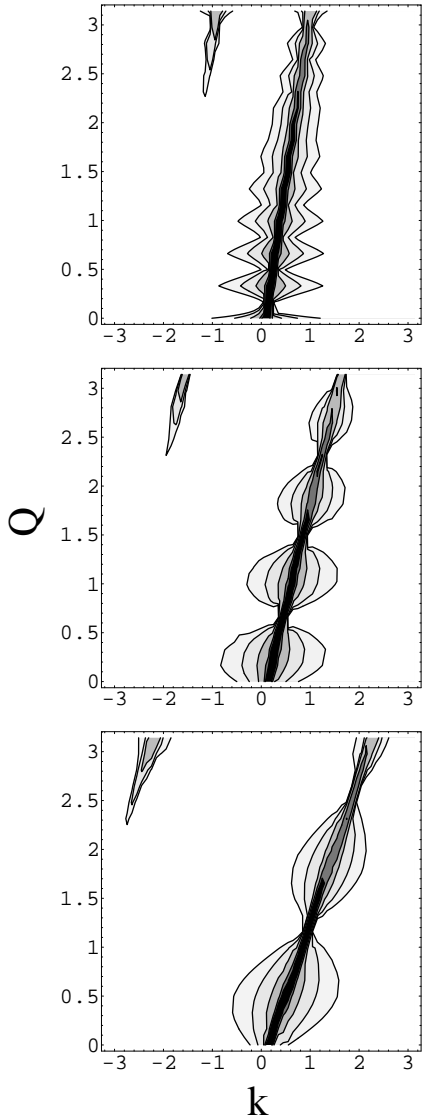}}
\caption{
The absolute value of  $A(Q,k)$ defined in (\ref{fou-def}) plotted
for the filling ratios $c=0.3, 0.5$ and 0.7, from top to bottom,
 respectively. The contours
are drawn at 0.03, 0.05, 0.1, 0.3 and 0.5.  The appearance of umklapp
solution (\ref{flaw}) is seen at large $Q$.
\label{fig:spectrum}}
\end{figure}

It is instructive to analyze at this step
the Fourier spectrum of the above amplitude :

     \begin{eqnarray}
     \label{fou-def}
     A(Q,k) =  \sum _{l=1}^L e^{-ikl}  A(Q,l)
     \end{eqnarray}
Again if our initial guess about the dependence
(\ref{defc}) in the whole range of $Q$
were true, we  would observe the following simple behavior
     \begin{eqnarray}
     \label{true}
     A^{(1)}(Q,k) \propto \delta(k-cQ)
     \end{eqnarray}
One can see on the contour map of   $A(Q,k)$ (Fig. \ref{fig:spectrum})
that at small $Q$
our initial expectations are satisfied. At the same time at larger $Q$
the second solution comes into play
     \begin{eqnarray}
     \label{flaw}
     A^{(2)}(Q,k) \propto  \delta(k-c(Q-2\pi))
     \end{eqnarray}

This finding shed some light on the nature of the discussed discrepancy.
Really the ``umklapp'' term $ A^{(2)}$ may take its origin in the
multi-sheet structure of the above logarithm (\ref{logarithm}) on
the complex plane.

At $Q\gtrsim 1 $ when the series  (\ref{logarithm}) diverges, the
value of $\ln f_q = n_q \ln e^{iQ}$ could be both $iQ n_q$ and
$i(Q-2\pi) n_q$, i.e. the value of logarithm
{\em on the other sheet} of complex
plane.

We conjectured that the correct answer
may be a linear
combination of the bosonized solutions ${\cal C}_{boson}(Q,l)$ (\ref{defc})
and  ${\cal C}_{boson}(Q-2\pi,l)$. We tried to fit
therefore the $l-$dependence of ${\cal C}(Q,l)$ at fixed $c$ and $Q$ :

     \begin{equation}
     {\cal C}(Q,l)_{fit}  =
     \alpha_1(Q) {\cal C}_{boson}(Q,l)
       + \alpha_2(Q) {\cal C}_{boson}(Q-2\pi,l)
     \label{A-fit}
     \end{equation}
with the real functions $\alpha_{1,2}(Q)$ as the parameters of the fit.

The  results of this fit are encouraging. The dependence  (\ref{A-fit})
and the calculated values (\ref{toeplitz})
are {\em indistinguishable} at $l\gtrsim 1$, $0\leq Q\leq \pi$ and
at $0.2\leq c \leq 0.8$.

\begin{figure}
\centerline{\epsfxsize=8cm \epsfbox{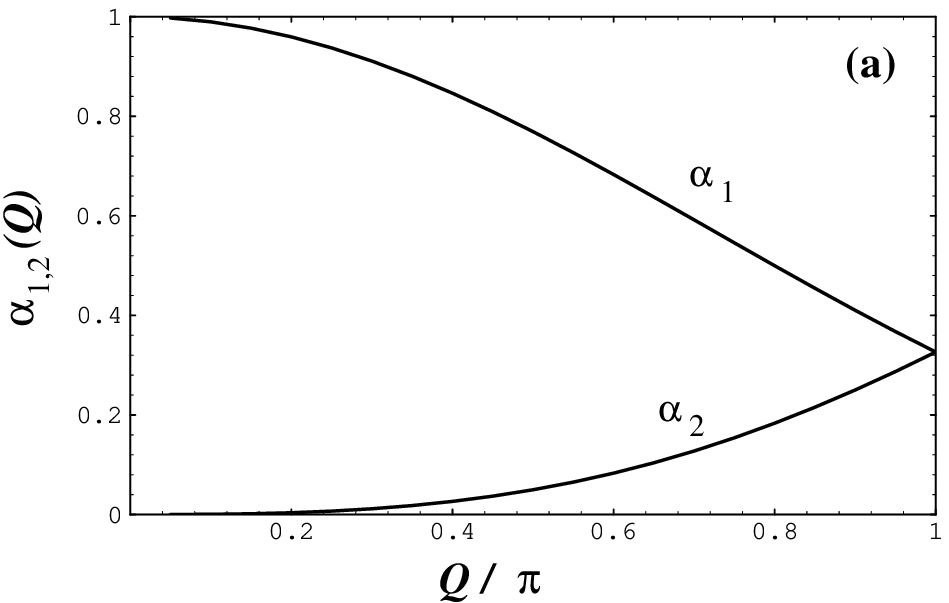}}
\centerline{\epsfxsize=8cm \epsfbox{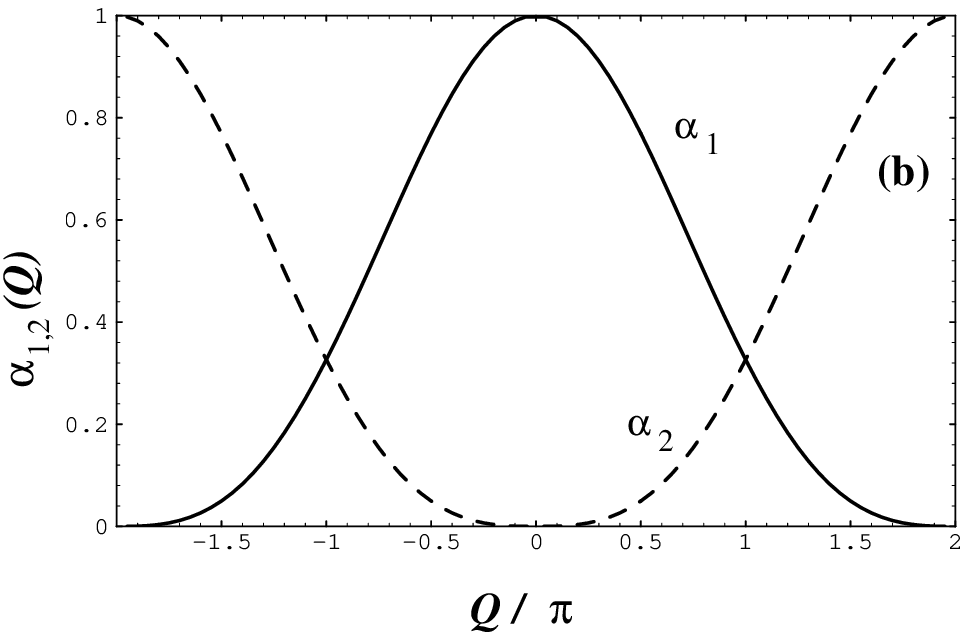}}
\caption{
a) The amplitudes of the  solutions
 $\alpha_{1,2}(Q)$ as determined by fitting
the calculated quantity $A(Q,l)$
with the Eq.\ (\ref{A-fit}) at $c = 0.5$.
b) The same functions shown in the extended range of $Q$.
\label{fig:alpha}}
\end{figure}

The above functions $\alpha_{1,2}(Q)$ slightly vary with $c$ in the above
interval, and are symmetrical with respect to the point $c=0.5$ as
one could expect in view of the particle-hole symmetry of the
initial Hamiltonian. The typical variance in $Q$ is shown on the
Fig. \ref{fig:alpha}a
for $c=0.5$.

One sees  that $\alpha_1(Q=0)=1$ and $\alpha_2(Q=0)=0$,
while  the ending values  $\alpha_{1}(\pi)$
and $\alpha_{2}(\pi)$ coincide so as
to produce the real values of ${\cal C}(Q=\pi,l)$.
We noticed next that the slopes of $\alpha_{1,2}(Q)$ at $Q=\pi$
coincide differing in sign.  We redraw
the obtained picture by shifting it by $\pm 2\pi$ as shown on Fig.
\ref{fig:alpha}b. Recalling the  $2\pi$-periodicity of the initial expression
(\ref{spinop1}) it becomes evident that actually $\alpha_{1}(Q)$
and $\alpha_{2}(Q)$ are the same even function $\alpha(Q)$,
obeying the relation :

     \begin{eqnarray}
     \label{defalpha}
      \alpha(Q)  &=&
     \alpha_{1}(Q) ,  \quad Q < \pi
      \\
     &=& \alpha_{2}(2\pi-Q), \quad Q > \pi \nonumber
     \end{eqnarray}

We now come to our main hypothesis. In view of the multisheet
structure of the logarithm found in the expression for the asymptotic
behavior of Toeplitz determinant, we conjecture
that the asymptotic behavior of the
average (\ref{spinop1}) can be represented in the
following form

      \begin{equation}
      \label{mainhypo}
      {\cal C}(Q,l) = \sum_{m=-\infty}^\infty
       {\cal C}_{boson}(Q - 2\pi m,l) \alpha(Q-2\pi m)
      \end{equation}
We notice that, numerically, the function $\alpha(Q)$ is close to
Gaussian,

      \[
      \alpha(Q) \simeq \exp \frac{Q^2}{2\pi^2} \ln l_0,
      \]
with $l_0 <1$. It allows one to further compact the expression
(\ref{mainhypo}).  Using the definition of elliptic theta-function
$\vartheta_3(z,q) $ \cite{GR}, we write an approximate equality
(cf. Eq.(\ref{defc})) :

      \begin{equation}
      \label{approx}
      {\cal C}(Q,l) \simeq \exp \left[iQcl -
      \frac{Q^2}{2\pi^2} \ln \frac l{l_0}  \right]
      \vartheta_3\left(\pi cl +\frac{iQ}\pi \ln \frac l{l_0},
      \frac {l_0^2}{l^2} \right)
      \end{equation}
which holds for the arbitrary $Q$.

Let us discuss here the particular consequences of Eq.(\ref{mainhypo})
in view of our previous findings. \cite{AG}.  Using the bosonization
approach we determined the {\em dynamics} of the fermionic correlations
on the partially filled chain; this result could hardly be obtained by
another technique. Although we were provided only by the bosonized
solution of the type (\ref{defc}), our main conclusions in \cite{AG}
preserve as explained below.

Let us write the bosonized solution in the form
${\cal C}_{boson}(Q,l )=  e^{iQcl} F(Q,l) $, the Fourier transform
$ F(Q,k)\sim |k|^{Q^2/2\pi^2-1}$. Then the quantity entering the formula
(\ref{sqw}) takes the form

      \begin{eqnarray}
      {\cal C}(Q,Q-k) &=&
      \sum_{m,n}\int\frac{dq}{2\pi}
      F(Q-2\pi m,q) \alpha(Q-2\pi m)
      \nonumber \\ \label{full} && \times
      \delta[(Q-2\pi m)(1+c)- k-2\pi n +q].
      \end{eqnarray}
The terms in (\ref{mainhypo}) with $m\neq0$
stemming presumably from backscattering in real space are
manifested in (\ref{full}) as the umklapp processes on the fictitious
lattice. They restore the invariance of the observable quantity
$\chi(k)$ (\ref{sqw}) upon the shift $Q\to Q+2\pi$, i.e.\
upon the choice of the Brillouin zone on the fictitious lattice.
This corresponds to the assumed cyclic boundary conditions.

It was shown in \cite{AG} that $\chi(k)$ had a principal
contribution from $Q\simeq \pm\pi$, $q\simeq0$. From the form of
$\alpha(Q)$ and the restriction $|Q|<\pi$ it is clear that only
the terms $m=0,\pm1$ are present in (\ref{full}).  A simple
analysis shows then that the only modification of our previous
results would be a symmetrization of the left and right
shoulders of the function $\chi(k)$ (\ref{sqw}) around its peak
values.  (see Fig.\ 3 in Ref.\  \cite{AG})

\section{discussion}

Firstly one notices that (\ref{mainhypo}) is explicitly $2\pi$ periodic
in $Q$.  In view of evenness of $\alpha(Q)$ the Eq. (\ref{mainhypo}) is
real at $Q=\pi$ as it must.

Secondly, the leading asymptotic behavior
at $|Q| < \pi $ is obviously delivered by $m=0$ term in (\ref{mainhypo}),
which is the term produced  by the bosonization procedure.
At the same time, at $|Q| \gtrsim 1$ the divergence of the underlying
logarithm series (\ref{logarithm}) manifests itself in the appearance
of the ``neighboring'' in $Q$ solution. This second solution is
relatively unimportant at
\[
\exp\left( -\frac{(2\pi - Q)^2 - Q^2 }{ 2\pi^2} \ln l \right)
< \exp ({-1})
\]
therefore one can ignore it only
 for exponentially large distances
\begin{equation}
l > \exp[{\pi}/{2(\pi - Q )} ].
\end{equation}
This argument can be also applied to the Fig.\
\ref{fig:spectrum} where the second
component is present mostly at
$\pi- Q \lesssim \pi/(2 \ln L) \simeq 0.4$

Thirdly, the bosonization is usually expected to work well, if the
filling ratio $c$ is close but not equal to $0.5$. This is to preserve
the almost linear character of dispersion near the Fermi level and to
avoid the influence of $2k_F$ processes. We saw that the Eqs.
(\ref{mainhypo}), (\ref{approx}) held for a rather wide range of $c$,
hence $2k_F$ processes were important for all $c$, while the bosonized
solution was a robust core in the final expressions.

Fourth, the algebra of the operators $b_q$,
$b^\da_q$, $N_\pm (U_\pm)$ is complete \cite{Haldane} and a proper
treatment should produce consistent expressions at any $Q$. This could
be an unfeasible task in view of the complexity of expressions
sketched in Appendix \ref{app:2kf}. Here we note that since
(\ref{mainhypo}) represents the asymptotic behavior, $\alpha(Q)$
could be non-analytic function. This point and the very structure of
Eq. (\ref{mainhypo}) implies that the thorough consideration of
backscattering terms in the fermionic density operator can lead to
(\ref{mainhypo}) only by careful resumming of the intrinsic divergences
in intermediate expressions.

Thus we observe that Eq. (\ref{mainhypo}) does not contradict to the
previously known results while its detailed analytical verification
may demand enormous efforts.

At the moment it is unclear whether our result should be regarded as a
very particular one or it could be found in other models, too.
The power-law decay of correlations is usually analyzed in terms of
leading asymptote, the existence of faster decaying one, as it is in
our case, may be missed. This should be compared to the results of Ref.  
\cite{GiSch}, where the faster decaying asymptotes of correlations were 
analyzed. It was shown, in particular, that these next-to-leading terms 
may dominate for the models with finite-range fermionic interactions. 

In conclusion we show that the Wigner-Jordan-like
exponential form in fermionic density operators reveals singularity
at some finite value of ruling parameter. The appearing additional terms,
apparently stemming from the $2k_F$ processes on real lattice,
correspond to umklapp processes on the fictitious lattice.
These terms restore a certain symmetry of the problem, lost after
the simple use of the bosonization technique.

\acknowledgements
We benefited from discussions with S. Aubry, H.B. Braun, G.S.
Danilov, A. Luther.  This work was supported in part by the RFBR
Grant No.\ 96-02-18037-a,  Russian State Program for
Statistical Physics (Grant VIII-2) and
Russian Program "Neutron Studies of the Condensed Matter".

\appendix

\section{the connection between fictitious and real lattices}
\label{app:connec}

Let us consider a chain with $2N$ sites and $N(1+c)$ electrons on it,
we assume $N\gg1$, $0<c<1$. The sites labeled by integer index are always
occupied by electrons. The remaining $cN$ electrons reside on $N$ sites
with half-integer indices. No double occupancy is allowed. The elementary
kinetic process (disregarding the spins) is the movement of electrons
on the half-integer sites. This process can be thought as combined of
two steps, first one is to allow the electron from the site $i$ to hop
onto the neighboring (empty) half-integer site $i+1/2$. At the second step
an electron moves from $i-1/2$ to $i$. One sees that upon this processs
the spin sequence of the whole chain preserves. One can also adopt that
the magnetic exchange between all neighboring spins has the same value.

 For such a system,
it is convenient to introduce the notion of fictitious lattice of spins
with $(1+c)N$ sites. The spin dynamics {\em on the fictitious lattice} is
then described by the Heisenberg Hamiltonian
       \begin{eqnarray}
       \label{h2}
       H_2 &=& J \sum_{i=1}^{N+cN-1}
        {\bbox \sigma}_i . {\bbox \sigma}_{i+1},
       \end{eqnarray}
while the kinetics of the system is that of the fermionic gas
(\ref{hamilt}).

Despite of the simple form of $H = H_1+H_2$ the observable
spin-spin correlation function,

\begin{eqnarray}
\label{sqwdef}
\chi(k)=
\left\langle {\bf S}_k.{\bf S}_{-k}\right\rangle,
\end{eqnarray}

\noindent is non-trivial because the relationship between
${\bf S}_i$ and ${\bbox \sigma}_i$ is not simple \cite{Vi-Bak}. The
difference between the coordinates of a integer-labeled spin measured in
the real and in the fictitious lattices is given by the number
of fermions located to its left. In terms of ${\bbox \sigma}_k$, the
Fourier transform of the spins on the fictitious lattice,  we
have:

\begin{eqnarray}
\label{spinop2}
{\bf S}_l=\frac1{\sqrt{N+cN}} \sum_Q^{} {\bbox \sigma}_Q \exp\ i Q
(l+\sum_{j=1}^{l-1} c_{j+1/2}^{\da} c_{j+1/2}^{}),
\end{eqnarray}

\noindent Substituting Eq.\ \ref{spinop2}  in
Eq.\ \ref{sqwdef} and using the fact that the ground-state
wavefunction of the decoupled Hamiltonian factorizes,
$\chi (k)$ may be written as a convolution  \cite{fnote2} :

\begin{eqnarray}
\label{sqw}
\chi(k)=\int_{-\pi}^{\pi} \frac{dQ}{2\pi} f(Q)\  {\cal
C} (Q,Q-k),
\end{eqnarray}

\noindent The first factor in Eq.\ \ref{sqw}, $f(Q)$, is
the  spin correlation function of the fictitious
1D Heisenberg model,  \cite{MTBB}

\begin{eqnarray}
\label{heisen}
f(Q)=\left\langle
{\bbox \sigma}_{Q}.{\bbox \sigma}_{-Q}\right\rangle_{H_2}.
\end{eqnarray}

\noindent The second factor, ${\cal C}(Q,k)$, contains the
kinematic effects and is given by the  Fourier
transform of (\ref{spinop1}), where one can
 drop the factor $1/2$ labeling spinless fermion position.
Thus we see that the observable spin correlations $\chi(k)$ on
the real lattice are connected to those on the fictitious
lattice $f(Q)$ via the integral transform (\ref{sqw}) with the
kernel ${\cal C}(Q,k)$.
The principal contribution to (\ref{sqw}) is delivered by
$Q\simeq \pm \pi$ where $f(Q)$ logarithmically diverges
\cite{AG}.

\section{temperature effects}
\label{app:temp}

At finite temperature we calculate the average of the form (\ref{defb})
via the basic identity :

      \begin{equation}
      \langle A\rangle =  {\rm Tr}\left(e^{-\beta H} A \right)
      /{\rm Tr}\left(e^{-\beta H} \right)
      \end{equation}
The  eigenstates of the bosonized Hamiltonian are given by \cite{Haldane} :
      \begin{equation}
      |N_+,N_-,\{n_q\} \rangle = (U_+)^{N_+} (U_-)^{N_-}
      \prod_{q\neq 0} \left(\frac{(b^\da_q)^{n_q}}
      {(n_q!)^{1/2}} \right)
       |0\rangle
      \end{equation}
Here $U_+$ ($U_-$) is the ladder operator which raises the fermion
charge $N_+$ ($N_-$) in unit steps and commutes with bosons $b_q$.

It is then straightforward to show that the temperature-dependent
correlator is given by the expression :
      \begin{equation}
      {\cal C}(Q,l ) = {\cal C}(Q,l )_{T=0} F(\beta=1/T,l)
      \end{equation}
with
      \begin{eqnarray}
       F(\beta,l) &=& \exp\left[ - Q^2 \sum_{q}
         \frac{|v(q)\phi(q,l)|^2}{\exp(\beta|q|v_F)-1} \right]
       \nonumber \\ && \times
       \frac{\sum_n\exp[-(\beta\pi v_F n^2-iQln)/N] }
        {   \sum_n\exp[-\beta\pi v_F n^2/N]}
      \label{temp}
      \end{eqnarray}
It is convenient to introduce here the temperature correlation length,
$\xi  = \beta v_F$.
The first factor in (\ref{temp}) stems from the boson contribution
and takes the form
\[
\exp -\frac {Q^2}{\pi^2} \ln\frac{\sinh\pi l/\xi}
     {\pi l/\xi} \sim \exp-\frac {Q^2l}{\pi\xi}
\]
in accordance with the result in \cite{AG} found with the use
of Szeg\"o-Kac theorem.

The second factor  in (\ref{temp}) stems from the charge fluctuations
and is relatively unimportant for all temperatures.
If $\xi \gg N$ then this second factor is unity and
the temperature does not break the coherence within the length
of a chain (or, more precisely, on a ring in view of implicitly assumed
periodic boundary conditions).
In the opposite case $\xi \ll N$ we have the
Gaussian decay of correlations :
$\exp [-Q^2 l^2/(4\pi \xi N)]$
The appearing  new length scale $\sqrt{\xi N}\gg \xi$ is however
irrelevant, since the correlations beyond $\xi$ are already
damped due to the boson contribution.

\section{``backscattering'' terms}
\label{app:2kf}

The consideration of the low-energy fluctuations of the electronic
density $\rho(q)$ with $q\simeq 2k_F$ produces the extra term to the
expression (\ref{defrho})
      \begin{eqnarray}
      c_{l}^{\da} c_{l}^{} &=&
       \sum_q e^{iql}[\rho_{+}(q)+\rho_{-}(q) ]
      + \tau(l) \\
      \tau(l) &\sim& \left(
      e^{i\Phi_J^{\da}(l)} U_+ U_-^{\da} e^{i\Phi_J(l)}
       + h.c.\right)
      \end{eqnarray}
with the ``current'' variable \cite{Haldane}
      \begin{eqnarray}
      \Phi_J(l) &=& k_Fl + \frac{\pi l}N (N_++N_-)
       \\ &&  \nonumber
      + i\sum_q \sqrt{\frac{2\pi}{N|q|}} sgn(q)e^{-iql} b_q
      \end{eqnarray}

It is seen that ${\cal C}(Q,l )$ contains now the exponential
from the linear combination and the exponential in bose operators.
Presently we did not succeed in evaluating of this type of expression.

\begin{references}
\bibitem{fnote} The fact of non-zero anticommutator $\{A_i, B_i\}$
does not change the Eq. (\ref{wick}), though it demands more
sophisticated arguments. \cite{AGD}
\bibitem{fnote2} For clarity we omit here the coherence factor
arising from consideration of half-integer sites.
It was shown in \cite{AG} that this factor produces some additional
modulation of the observable quantity $\chi(k)$.
\end{references}
\end{document}